\documentstyle[11pt,newpasp,twoside,epsf]{article}
\begin{document}
\title{Stars and Planetary Nebulae in the Galactic Bulge}
\author{F. Cuisinier}
\affil{Depto. de Astronomia, UFRJ,
       Rio~de~Janeiro, Brazil}
\author{J. K\"oppen \& A. Acker}
\affil{Observatoire de Strasbourg, 
        France}
\author{W.J. Maciel}
\affil{Instituto Astron\^omico e Geofisico da USP, 
       S\~ao~Paulo, Brazil}

\section{Introduction}

Planetary Nebulae are interesting objects for the study of the Galactic Bulge 
because they concentrate the energy of their central stars in the emission
lines of their spectra, and can therefore be observed relatively easily at  
this distance. Furthermore the masses of their progenitor stars varying 
from 0.8
to  ${\rm 8 M_{\sun}}$,  their ages span from 50 Myr to 25 Gyr,
 covering  more 
than 95\% of the possible ages in the Universe, and of course 
in the Bulge.

The only stars with a similar ages range, that are reasonabely
 observable in the 
Bulge,  are the Red Giants. They  are actually the direct precursors of
the Planetary Nebulae. Some elements have their abundances unmodified
by the stellar evolution in Red Giants as well as in Planetary Nebulae.
These elements keep the fingerprints of the chemical composition of the
ISM when the progenitor star was born, and because of the span of their ages, 
they allow to follow its evolution over a very wide time range.

One particular point of interest are the relative abundances of elements
produced in type II and in type Ia supernovae. Type II supernovae explode
very rapidely, after some Myr, e.g. quasi instantaneously on 
the Bulge evolution timescale, whereas type Ia supernovae explode after
a period of the order of one  Gyr. 
The relative abundances of type II and type Ia supernovae should thus 
allow to measure the timescale of the Bulge formation.

On the other hand, elements produced during the lifetimes of the progenitor
stars should allow to determine their ages - at least statistically. 
In Planetary Nebulae, nitrogen is very easily detectable, and has its 
abundance modified in high mass progenitors, that are short lived.
Nitrogen abundances in Planetary Nebulae should thus  
help to identify recent star formation.

\section{Abundances in Stars and in Planetary Nebulae}
 
We derived  abundances for a sample of 30 PN, that we observed with 
high quality spectroscopy (Cuisinier et al. 2000). These abundances
being of really better quality than others available in the 
literature, we will only consider these ones here.

Abundances for individual elements in stars are up to now only available
for a sample of 11 red giants, from McWilliam \& Rich (1994).

Unfortunately,  a direct comparison of  abundances is  not possible,
the elements detectable in stars with a good confidence 
being different
from those detectable in Planetary Nebulae. 

We compared therefore the distributions of O, S and Ar in Planetary Nebulae
in the Bulge and in the Disk, these elements representing the pristine 
abundances of the ISM (Figure 1, left panel, for O). We found  the 
abundances distributions to be quite similar, like the Fe abundances
in the stars (Mc William \&  Rich 1994). 

On the other hand, the N/O ratios comparison in the Bulge and the Disk
(figure 1, left panel) show that the young progenitor, N-rich Planetary 
Nebulae, that are present in the Disk, are lacking in the Bulge. From the 
Planetary Nebulae, the Bulge does not seem to have formed stars 
recently.\\

If the Red Giant and the Planetary Nebulae populations in the
Bulge seem to be quite similar  in the light of our study, the picture
that arises from a comparison of the various  elements originating from
type II and type Ia supernovae that are
detected in Planetary
Nebulae and in Red Giants  remains very puzzling: Mg and Ti, that are enhanced
over Fe, 
seem to favor a quick evolution, whereas He, O, Si, S, Ar and Ca show
normal abundances patterns, and favor  a much slower evolution.

\begin{figure}
\plottwo{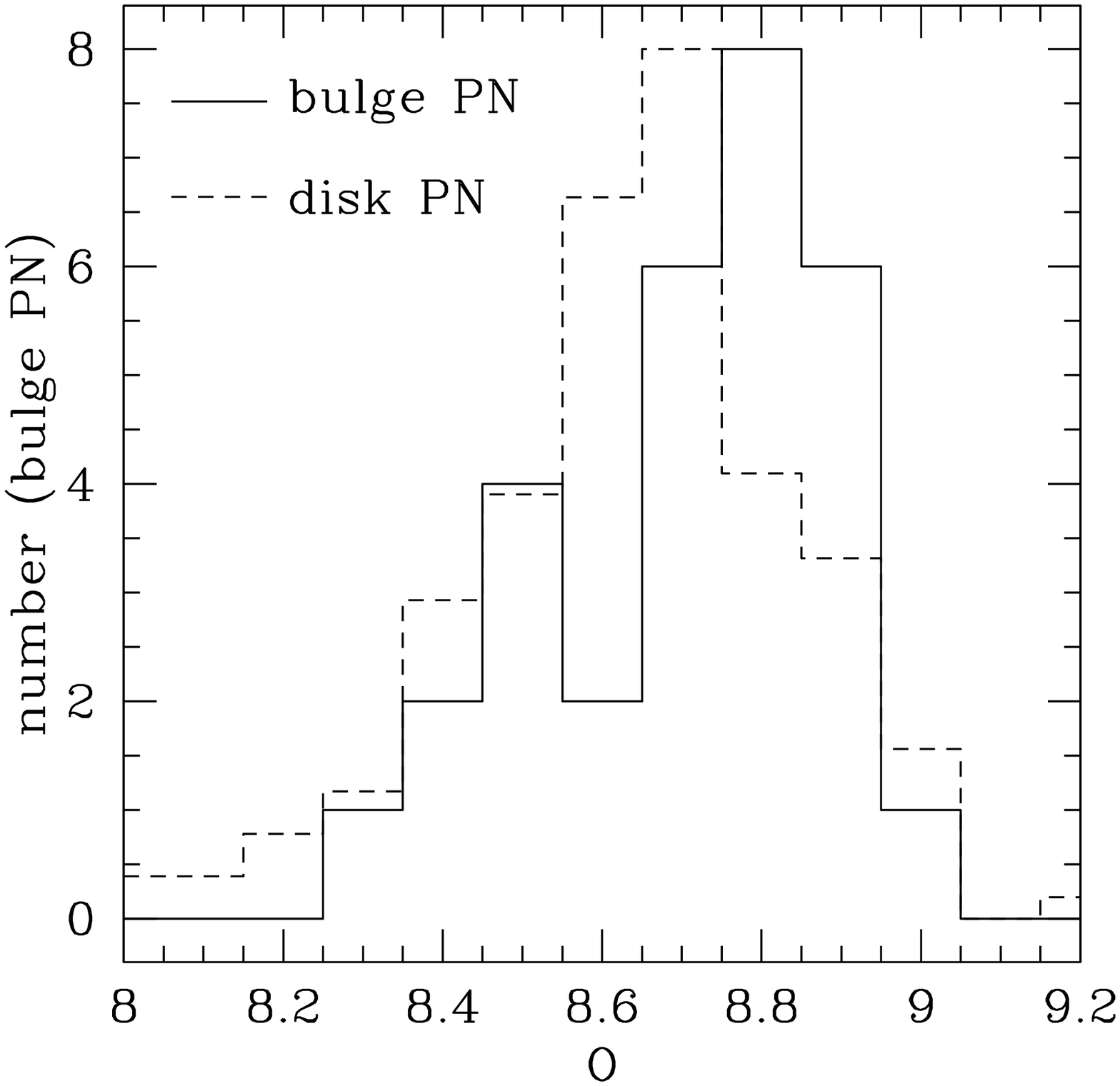}{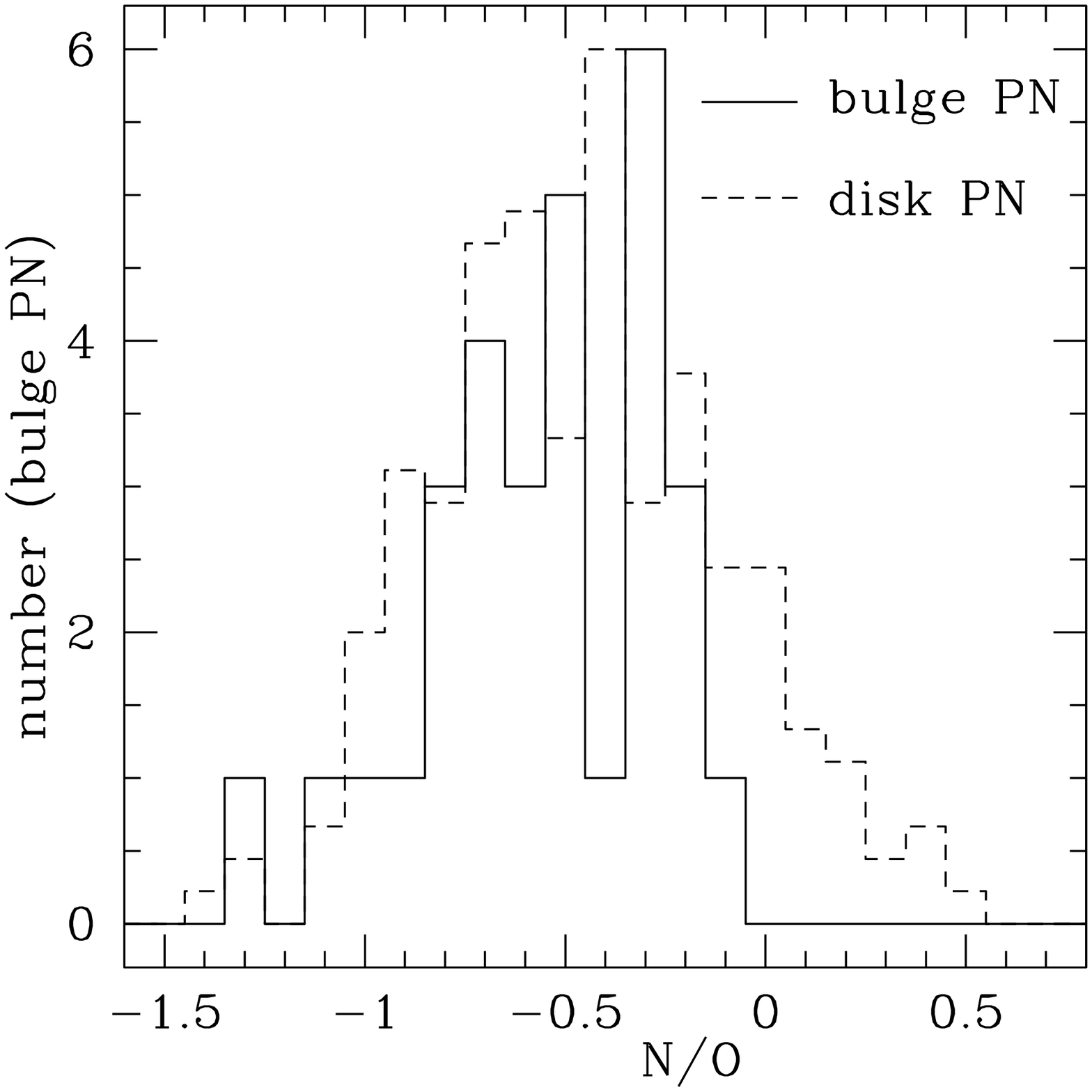}
\caption{Left: O abundances in PN. Right: 
N/O ratios in PN}
\end{figure}


\begin{references}

\reference Cuisinier, F., Maciel, W.J., K\"oppen, J., Acker, A., 
           Stenholm, B. 2000, \aap, 353, 543

\reference McWilliam, A., Rich, R.M., 1994, \apjs, 91, 749

\end{references}
\end{document}